%% file: steerage.tex
\documentstyle[12pt]{article}
\def\beq{\begin{equation}}
\def\eeq{\end{equation}}

\begin{document}

\vspace*{10mm}
\noindent {\bf PILOT WAVE STEERAGE: A MECHANISM AND TEST }\\[3mm]
\hspace*{15mm} Found. Phys. Lett. {\bf 12} (5), 441 (99)\\[7mm]
\hspace*{15mm} A. F. Kracklauer \\[5mm]
\hspace*{15mm} {\it Belvederer Allee 23c}
\\
\hspace*{15mm} {\it 99425 Weimar, Germany}
\\
\hspace*{15mm} {\it kracklau@fossi.uni-weimar.de}\\[5mm]
\hspace*{15mm} Recieved 6 April 1998; revised 29 August 1999\\[5mm]

\noindent
An intuitive, generic, physical model, or conceptual paradigm for pilot 
wave steerage of particle beams based on Stochastic Electrodynamics is 
presented.  The utility of this model for understanding the Pauli 
Exclusion Principle is briefly considered, and a possible experimental 
verification for the underlying concepts is proposed.  \\[7mm] Key 
words: Quantum Mechanics, Pilot Wave, Pauli Exclusion Principle, 
Stochastic Electrodynamics \\[7mm] 

\bigskip\noindent {\bf 1. PROBLEM DEFINITION AND BACKGROUND}\bigskip

Feynman said of the diffraction of particle beams: ``In reality, it 
contains the only mystery'' [in Quantum Mechanics].~$^{(1)}$  This 
is so for lack of an explanation of how individual particles  `know' 
about the geometry of the objects which  cause beam diffraction; for 
example, whether there is one or two slits in Young's experiment.
Two paradigms dominate theorizing on
this question. The prevailing 
orthodox view is `dualism,' or
`complementarity,' which holds that while in transit, the `wave' 
nature `feels' the boundaries and determines behavior, but that at the 
instant of measurement, the wave
collapses to its complementary `particle' nature.   
The main, perhaps only, alternatives are variations on de Broglie's 
pilot wave notion.  Historically, de Broglie's core idea was that the 
ontological essence of a particle is in fact an object consisting of a 
particulate  kernel embedded in a wave which serves as a scout, 
guiding the kernel. 

As they are, neither of these concepts is natural.  The orthodox idea 
suffers profound problems for lack of a fundamental distinction 
between those interactions which are `measurements' and therefore cause 
collapse, and those that do not.  This is a deep problem in view of the 
fact that most measurements are made by capturing radiation, which at 
the moment of emission, could just as well never end in a 
measurement,  
astronomical observations, say.  The pilot wave theory, on the other 
hand, lacks a plausible mechanism for describing just how the wave 
arises and does its guiding.  All obvious explanations, to the extent 
any has been proposed,  lead to the expectation of high particle 
densities where the wave has nodes, the opposite of what is 
observed.~$^{(2)}$ 

This work is a contribution to the theory of pilot wave guidance.  
Its goal is only to cobble together a paradigm of components from 
classical physics to rationalize this element of Quantum Mechanics 
(QM), not to further analyze the foundations for deeper consequences 
of these components.  Such studies are left for the future; in the 
first instance, any classical rationalization of QM is by itself a 
breakthrough. 

As an aside at this point, note that de Broglie's pilot wave theory was 
inspiration for what has become known as the de Broglie-Bohm alternate 
interpretation.~$^{(3)}$ In this Bohm variant, the `scouting' 
function is attributed to an additional `quantum' potential (in some 
formulations, implicitly) for which the theory offers no further 
motivation, in particular, none with respect to classical wave 
phenomenon.  With this in mind, it therefore seems that Bohm's Mechanics 
is for the purposes of ontology equivalent to `Copenhagen' QM.  
Certain other alternate interpretations; e.g., Consistent-Histories, 
Many-Worlds and others, also seek to
invest QM with an 
internally consistent interpretation without reference to wave or 
other concepts from classical physics.  As such, they belong to 
intellectually separate streams for which this study has no relevance, 
and of which it offers no evaluation, rather just competition on the 
field of intuitive appeal.     

\bigskip\noindent {\bf 2. FUNDAMENTAL CONCEPTS}\bigskip

The foundation of the model or conceptual paradigm for the mechanism 
of particle diffraction proposed herein is Stochastic Electrodynamics 
(SED).   Most of SED, for which there exists a substantial literature, 
is not crucial for the issue at hand.~$^{(4)}$  The {\it nux} of 
SED can be characterized as the logical inversion of 
QM in the following sense.  If QM is taken as a valid theory, then 
ultimately one concludes that there exists a finite ground state for 
the free electromagnetic field with energy per mode given by 
$\hbar\,\omega/2$. SED, on the other hand, inverts this logic and 
axiomatically posits the existence of a random electromagnetic 
background field with this same spectral energy distribution, and then 
endeavors to show that ultimately, a consequence of the existence of 
such a background is that physical systems exhibit the behavior 
otherwise codified by QM.   The motivation for SED proponents is to 
find an intuitive local realistic interpretation for QM, hopefully to 
resolve the well known philosophical and lexical problems as well as 
to inspire new attacks on other problems. 

The question of the origin of this electromagnetic background is, of 
course, fundamental.  In the historical development of SED, its 
existence has been posited as an operational hypothesis whose 
justification rests {\it a posteriori} on results.  Nevertheless, 
lurking on the fringes from the beginning, has been the idea that this 
background is the result of self consistent interaction; i.e., the 
background arises out of interactions from all other electromagnetic 
charges in the universe.~$^{(5)}$ 

For  present purposes, all that is needed is the {\it Ansatz} that 
particles, as systems with charge structure (not necessarily with a 
net charge), are in equilibrium with electromagnetic signals in the 
background.  Consider, for example, as a prototype system a dipole 
with characteristic frequency $\omega_{0}$.  Equilibrium for such a 
system can be expressed as \beq m_{0}c^{2}=\hbar \omega _{0}. \eeq 
This statement is actually tautological, as it just defines $\omega_0$ 
for which an exact numerical value will turn out to be practically 
immaterial. 

This equilibrium in each degree of freedom is achieved in the 
particle's rest frame by interaction with counterpropagating 
electromagnetic background signals in both polarization modes 
separately which, on the average, add to give a standing wave with 
antinode at the particle's position: 
\beq 2\!\cos\left(k_{0}x\right)\sin\left(\omega _{0}t\right). \eeq 
Again, this is essentially a tautological statement as a particle 
doesn't `see' signals with nodes at its location, thereby leaving only 
the others.  Of course, everything is to
be understood in an on-the-average, statistical sense.     

Now consider Eq. (2) in a translating frame, in particular the rest 
frame of a slit through which the particle as a member of a beam 
ensemble passes.  In such a frame the component signals under a
Lorentz transform are Doppler shifted and then add together to give 
what appears as modulated waves, 
\beq 2\cos\left(k_{0}\gamma \left(x\,-\,c\beta 
t\right)\right)\sin\left(\omega _{0}\gamma \left(t\,-\,c^{-1}\beta 
x\right)\right),  \eeq
for which the second, the modulation factor, has wave length $\lambda = 
\left(\gamma\, \beta k_{0}\right)^{-1}$. 
 
From the Lorentz transformation of Equ. (1),                           
\beq P'=\hbar \gamma \beta k_{0}, \eeq 
$\gamma\,\beta k_{0}$ can be identified as the de Broglie wave vector 
from QM as expressed in the slit frame. 

In short, it is seen that a particle's de Broglie wave is modulation on 
what the orthodox theory designates Zitterbewegung.  The modulation 
wave functions as a pilot wave.  Unlike de Broglie's original 
conception in which the pilot wave emanates from the kernel, here this 
pilot wave is a kinematic effect of the particle interacting with the 
SED Background.  Because this SED Background is classical 
electromagnetic radiation, it will diffract according to the usual 
laws of optics and thereafter, modify the trajectory of the particle 
with which it is in equilibrium.~$^{(6)}$ (See Ref. 4., Section 
12.3, for a didactical elaboration of these concepts.)

\bigskip\noindent {\bf 3.  PILOT WAVE STEERAGE}\bigskip

All the above is a brief review of concepts to be found in the 
literature, in part for up to 70 years.  What remains unanswered, 
however, is the question of just how a pilot wave steers a 
particle.  This question is made particularly vexing in that obvious 
mechanisms seem to lead to a close, but still wrong answer.  
Specifically, if it is imagined that particles are nudged by the 
radiation pressure of pilot waves, then particles should be found 
preferentially at the nodes of these waves where pressure is lowest.  
But this is not so.  Neutron diffraction experiments, and others, 
yield classical Fraunhofer single slit patterns with a distinct 
central hump---if radiation pressure from the  pilot wave is the 
steering mechanism, there should not be a central hump but twin humps 
located at the flanking nodes.~$^{(7)}$ Clearly, something is 
missing. 

It is the purpose herein to suggest additions to this model to amend 
this deficiency.   The basic concept exploited to achieve this end is 
to take the modulation function of  pilot waves seriously, and to 
observe that the energy pattern of the actual signal that  pilot waves 
are modulating, and to which a particle tunes, comprises a fence or 
rake-like structure with prongs of varying average heights specified 
by the pilot wave modulation.  These prongs in turn can be considered 
as forming the boundaries of energy wells in which particles are 
trapped.  Intuitively, it is clear that where such wells are deepest, 
particles will tend to be trapped and dwell the longest.  The exact 
mechanism moving and restraining 
particles is radiation pressure,  
but not as given by the modulation, rather by the carrier signal 
itself.  Of course, because these signals are stochastic, well 
boundaries are bobbing up and down somewhat so that any given particle 
with whatever energy it has will tend to migrate back and forth into 
neighboring cells as boundary fluctuations permit.  Where the wells 
are very shallow, however, particles are laterally (in a diffraction 
setup, say) unconstrained; they tend to vacate such regions, and 
therefore have a low probability of being found there.  

The observable consequences of the constraints imposed on the motion 
of particles is a microscopic effect which can be made manifest only 
in the observation of many similar systems. 
For illustration,  consider an ensemble of similar particles 
comprising a beam passing through a slit.  Let us assume that these 
particles are very close to equilibrium with the background, that is, 
that any effects due to the slit can be considered as slight 
perturbations on the systematic motion of the beam members.  

Given this assumption, each member  of the ensemble with index $n$, 
say, will with a certain probability have a given  amount of kinetic 
energy, $E_n$, associated with each degree of freedom.  Of special 
interest here is the beam direction perpendicular to both the beam 
and the slit in which, by virtue of the assumed state of near 
equilibrium with the background, we can take the distribution with 
respect to energy of the members of the ensemble to be given in the 
usual way by the Boltzmann Factor: $ e^{-\beta E_n}, $  where $\beta$ 
is the reciprocal  product of the Boltzmann Constant $k$ and the 
temperature, $T$, in degrees Kelvin.  The temperature in this case is 
that of the electromagnetic background serving as a thermal bath for 
the beam particles with which it is in near equilibrium.  

Now, the relative probability of finding any given particle; i.e., with 
energy $E_{n,j}$ or $E_{n,k}$ or $\dots$, trapped in a particular well 
will be, according to elementary probability,  proportional to the sum 
of the probabilities of finding particles with energy less than the 
well depth, $d$, say: 
\beq \sum_{\{l|E_{n,l} \le d\}} e^{-\beta E_{n,l}} \cong \int_0^d
d(E_n/{\cal E}_0)e^{-\beta{\cal E}_n} 
= \,\,{1 \over{\beta{\cal E}_0}}\left(1-e^{-\beta\,d}\right),  
\label{ansatz}  \eeq 
where approximating the sum with an integral is tantamount to the 
recognition that the number of energy levels, if not {\it a priori} 
continuous, is large with respect to the well depth.  

If now $d$ in this equation is expressed as a function of position, we 
get the probability density as a function of position.  For example,  
for a diffraction pattern from a single slit of width $a$ at distance 
$D$, the intensity (essentially the energy density) as a 
function of lateral position is: ${{{{{\cal 
E}_0 \sin^2 (\theta)}}}/{{\theta^2}}}$, where $\theta = k_{pilot 
wave}({2a\over{D}})y$, and the
probability of occurrence, $P(\theta(y))$, as a 
function of position, would be 
\beq P(y) \propto  \left(1 - e^{-\beta {\cal 
E}_0\,{\sin^2 (\theta)}\over{\theta^2}}\right). \label{prob}  \eeq  

Whenever the exponent in Eq. (\ref{prob}) is significantly 
less than one, its 
r.h.s. is  very accurately approximated by the exponent itself;
 so that one obtains the standard and verified 
result that the probability of occurrence,  $\psi^*\psi$ in 
conventional QM, is  proportional to the intensity 
of a particle's deBroglie (pilot) wave. (See Ref. 6. for an 
account relating $\psi^*\psi$ to a probability and $\psi$ to 
a pilot wave on the basis of SED.)

For more complex particles which have more than just a dipole 
interaction, the carrier wave becomes more complex.  In turn, the spike 
structure becomes more complex, but the general considerations above 
remain valid.  In any case, the spike structure is on a scale much 
finer (at Zitter frequencies and Compton-like wave lengths) than 
the modulation, and would therefore remain essentially unobservable so 
that only modulation patterns are manifest. 

The condition that the exponent in Eq.(\ref{prob}) is to be less 
than one, depends on contributions from two factors, $\beta$, and 
$d(y).$  The first of these reflects the
thermodynamic environment of 
the ensemble member in contact with the background as a heat bath.  
Elementary fundamental considerations set a limit on 
the term $\beta {\cal E}_0$.  If a particle trapped
 in an energy well as described 
above is regarded in its rest frame as exposed to an harmonic 
oscillator potential (as a first order
approximation to the energy well 
given by $sin^2({{m_0c}\over{\hbar}}y)$), then the mean total 
energy equals $kT$ while the mean kinetic energy is $kT/2$, and this 
implies that the exponent in Eq. (6) is less than $1/2.^{(8)}$ 

The second factor determining the magnitude of the exponent in Eq. 
(\ref{prob}) is the factor $d(y).$  Above, the expression used for 
single slit diffraction is the idealized Fraunhofer amplitude which 
ignores the $r^{-1}$ fall-off of the intensity.  In more accurate 
calculations, this fall-off contributes to a reduction of the exponent, 
thereby further improving the approximation.  

Exploitation of this deviation to experimentally verify the model would 
be probably very difficult.  Even in the limiting case when $\beta 
{\cal E}_0 = 1/2$, the deviation of, for example, a single slit diffraction 
pattern is slight.  Figure 1 compares the curve derived from Eq. 
(\ref{prob}) for a particle beam with that for radiation (i.e., the 
exponent in Eq. (\ref{prob})) where the curves are normalized so 
that their peaks coincide.  The geometry here mimics closely that of 
the single slit neutron diffraction experiments described in Ref. 
7.  The  deviation is startlingly small.                

\input neut.tex 
\dessin{Comparison of the single slit radiation diffraction pattern to 
that for a particle beam as given by Eq.
(\ref{prob}).  The slit geometry and
beam characteristics (i.e., background) 
correspond to those described in Ref. 7. for an experiment 
with neutrons.  The $\chi^2$-curve shows the contribution of the 
deviation in each displacement bin to
$\chi^2$ as used in regression 
analysis.  It is clear that an attempt to fit data described in nature 
by Eq. (\ref{prob}) with $sin^2(x)/x^2$ would fall well within the 
statistical significance of current experiments.  As the curves in 
this figure are computed with $\beta {\cal E}_0 = 1/2$, when in fact 
for neutrons this factor would be significantly less, the fit in fact 
is much better than shown here.}{Figure 1} 

Furthermore, if data described by Eq. (\ref{prob}) taken in a particle 
beam diffraction experiment is
 fitted using $\chi^2$-regression 
techniques to the radiation diffraction pattern, the fit can be 
improved by adjusting the assumed slit width.  When done, the result 
is approximately a $1.5\%$ reduction in the slit width.  It should be 
noted here that in Ref. 7. it was
reported that for neutron 
single slit diffraction, the fit to the pure radiation pattern was 
improved by assuming an approximately $6\%$ {\it increase} in the slit 
width. On the basis that an essentially identical result was observed  
for laser beam diffraction through the
identical optics chain, this 
was attributed to an artifact of the optical geometry of the 
experiment .  Thus, in combination, the effect discussed herein, 
seemingly would be observable only as a slight reduction of the 
increase and would be below the statistical significance of their 
experiments.  In particular, because the neutron is a complex particle 
for which $\beta {\cal E}_0$ can be expected to be less than $1/2$, 
the optimal slit size reduction would be less that $1.5\%$; e.g., for 
$\beta {\cal E}_0  = 1/5$, the reduction is $1\%$. (The curves in 
figure 1 address only the issue of the suitability of the radiation 
diffraction pattern for fitting data which is in fact described by Eq. 
(\ref{prob}) and not the specific geometry of the experiment described 
in Ref.~$^{(7)}$  for beam generation and measurement.) 

Electron diffraction patterns, as determined by the exigencies of 
experimental setups, are typically multi-peak patterns for which 
errors and tolerances overwhelm deviations attributable to the 
essential difference in particle beam and radiation diffraction 
patterns.~$^{(9)}$ See Figure 2. 

\input  electron.tex
\dessin{Comparison of particle beam and radiation multislit 
diffraction patterns corresponding to the experiment described in 
Ref. 9. for electron beams.  Here, although the value of $\beta {\cal 
E_0} = 1/2$ is fully appropriate, the deviation of the particle beam 
pattern from the radiation pattern is still well below the limit set 
by the statistical significance of current experiments.}{Figure 2}

Also of interest is the question regarding the coherence length of 
guiding waves.  According to standard theory, the coherence length of 
a signal, $\Delta l$, equals $c/ {\Delta \nu}$, where $c$ is the 
speed of light and $\Delta \nu$ is the bandwidth of the signal.  In 
this application, the bandwidth of the background is undefined as all 
frequencies are present in the background.  Nevertheless, the 
effective acceptance function of the particle,  arising, {\it inter 
alia}, as inverse line broadening from random motion  will result in 
the same thing. 

The lateral coherence area of background signals, also according to 
standard theory, is $\Delta A \sim R^2 {\lambda}^2/S$ where $R$ is the 
distance to the source, $\lambda$ is the wave length of the signal and 
$S$ is the surface area of the source.  Common astronomical sizes and 
distances attributed to specific sources of a particular background 
signal, e.g., $R \sim 10^9$ lightyears, $\lambda \sim 5\times 10^{-
12} m\, $(typical for electron beams), and $S~\sim~10^{-20}m^2\, 
$(atoms) to $S~\sim~10^8m^2\, $ (stars), lead to coherence widths 
circa $10^{2}$ to $10^{25} m$.  While these results must not 
be taken too seriously, they do confirm that straight forward 
estimates do not render the underlying concepts improbable. 

\bigskip\noindent {\bf 4.  PAULI EXCLUSION PRINCIPLE}\bigskip

In these considerations, the outline of a qualitative rationalization 
of the Pauli Exclusion Principle from an SED viewpoint may be 
emerging.   In SED, Spin can be seen as a manifestation of the  vector 
character of electromagnetic background signals.  In a magnetic, `$B$,' 
field, particle motion in a plane perpendicular to the this $B$ field 
can be resolved in terms of clock- and counterclockwise motion each 
separately in interaction with likewise polarized 
background-signals.~$^{(4, 6)}$  Such average circular motion 
gives rise to a magnetic dipole.  The energy difference between 
alignment and antialignment of these magnetic dipoles resulting from 
this background-driven gyration in typical laboratory $B$ fields is 
circa $ 10^{-8} $ that of the rest energy of the particle.  Thus, per 
the Boltzmann factor, the  populations in an ensemble of `spin up and 
down' particles are virtually equal when they effectively do not 
interact, for example, when they are distant, independent beam 
particles .  However, in an atom, where because of proximity strong 
interaction is inevitable, the energy difference between aligned and 
antialigned dipoles, being proportional to $ r^{-3} $ where $r$ 
is the separation, is large and  implies in turn that the 
equilibrium population distribution difference as given by the 
Boltzmann factor is large and in favor of the antialigned state.  In 
effect, aligned dipoles preferentially escape from the constraining 
wells envisioned above leaving only those antialigned states 
`permitted' by the Pauli Principle in the same diffraction pattern---
customarily denoted as a `quantum state.'  In an atom, of course, 
cyclicity, rather than geometrical boundaries such as slits 
determines standing wave patterns. 

In this paradigm, interaction is a mechanism to foster energetic 
differences in `states,' which then, according to the Boltzmann 
factor, result in population differences between these states. 
Likewise, monopole interaction, in this paradigm, would cause one or 
the other particle of a pair to experience energy excursions, the 
effect of which would cause it to exceed the retaining energy 
of the wells in which it is trapped.  Also,
with respect to dipole interaction, an essential conceptual element 
of this model is that  spin is engendered by a magnetic field which 
means, that because two electrons in close proximity (where $r^{-3}$ is 
large) are exposed to essentially the same magnetic field, the 
geometry of the spin interaction is
restricted to being parallel and 
antiparallel only.  This feature, as is well appreciated, 
distinguishes particles with `spin'  from classical dipoles. 

In summary, these considerations begin to render the Pauli Exclusion 
Principle intuitively consistent with classical thermodynamics given 
an SED background.  That it is a rigorous necessity in detail remains 
to be shown.

\bigskip\noindent {\bf 5. POSSIBLE EXPERIMENTAL TEST}\bigskip 

A possible test of the above concepts might be made by so arranging 
that both openings of a Young double slit experiment are transparent 
to pilot wave radiation, while only one of them is transparent to 
particles.  With electron beams this might be achieved, for example by 
applying a transverse electric field to slit A, say, while leaving 
slit B in its innate state.  If set up propitiously, particles passing 
through slit A will be forced away from the registration zone of the 
observation screen.   A particle passing through slit B, however, will 
remain in equilibrium with the double slit pattern as its pilot wave 
passes unaltered through both slits.  The consequent effect then, will 
be simply to reduce by half the intensity of the pattern seen on the 
observation screen. 

By way of contrast, if  the current orthodox interpretation of QM is 
correct, blocking the particles in slit A in any way should result in 
the interference pattern changing to that of a single slit as well as 
a reduction in the intensity.  A particle passing through a double 
slit is put into a `cat' state, $1/{\sqrt{2}}(|A\rangle + |B\rangle)$, 
which is then to interfere with itself to yield the double slit 
pattern.  If particles are prevented from passing through slit A with 
certainty, then the subsequent state can only have the $|B\rangle$ 
component, so that the wave function can exhibit only the single slit 
interference. 

As is usual with Young's double slit experiment on a microphysics scale, 
realizations are not unproblematic.   In this case, an additional 
crucial factor arises; namely, whatever is done to or in slit A, must 
not spill over into slit B and destroy the coherence of the beam 
passing through it by introducing dispersion in velocity. 
 Such spurious intervention, to first order at least, 
would destroy completely  the diffraction pattern rather than 
transform it from the double to single slit pattern. 

From the imagery afforded by the SED model of particle diffraction, it 
can be seen also that `which-way'  identifier operations in two-slit 
experiments that seek to tag particles passing one slit must do so 
such that phase shifts are not induced.  It is not sufficient that the 
input and output wave vectors of the tagging operation are identical.  
In order to test complementarity, it is also necessary that the 
tagging operation does not introduce a random phase shift with respect 
to that portion of the pilot wave that passes through the other slit.  
If such a phase shift is randomly distributed, ensemble uniformity is 
lost, diffraction patterns vanish, but principles remain untested. In 
particular, this means that tagging operations in which polarization 
is affected would be disallowed as the two modes are independent and 
therefore, the phases are randomly distributed from particle to 
particle even when no (net) work is done in the propagating direction. 

\bigskip\noindent {\bf 6.  ANCILLARY COMMENTS}\bigskip

In many particle beam experiments, the optical elements are not 
passive but actively introduce an intervention which is functionally 
equivalent to a measurement.  For electron beams, for example, two 
slits can be simulated with a so called biprism that consists of a 
charged wire parallel to a transverse direction of the 
beam.~$^{(10)}$ As the beam passes on each side of the wire, it is 
deflected away somewhat from the longitudinal direction of the beam so 
as to form two slightly diverging beams.  A second such wire parallel 
to, oppositely  charged and downstream from the first, then serves to 
draw the diverging half beams together again so that they converge and 
interfere on the observing screen.  In this arrangement the  beam 
particles (electrons) are deflected by work done by imposed electric 
fields and not by virtue of diffraction of matter waves (or by energy 
pattern wrinkling in SED induced pilot waves).  Since the beams, after 
passing a biprism can reconstitute  matter waves (i.e., reequilibrate   
with new signals in the background), experiments of this type 
seemingly can not reveal particle/pilot-wave feed-back or self-
interference, but rather interbeam interference. 
                                                
On the other hand, the fact that a biprism works at all, provides 
backhanded evidence that local hidden variables exist.  In 
conventional QM, the wave function is considered complete and uniform; 
there are no separate particle and wave aspects; the two qualities are 
totally intermingled.  Thus, when a single particle wave function is 
divided at a biprism, both the wave and particle aspects must be 
similarly divided.  However,  when a single particle wave is 
divided and measurements are made only on a portion of the beam, 
either nothing at all or the whole particle is observed.  Collapse of 
the wave packet at the moment of observation can be evoked to explain 
the appearance of the whole particle.  But this explanation runs amok  
when it is recalled that the division of the wave function in a 
biprism in the first place occurred by virtue of deliberate 
intervention, (i.e., by consciously evoked fields whose effect is 
recordable by observing the current in the prism wire---whether in 
fact done or not),  which is equivalent to a measurement.  Then, if 
the wave is collapsed at the prism, there should be no wave 
thereafter to interfere at the observing screen.  Moreover, if this
intervention is admitted into the class of agents provoking collapse, 
then, as these intervention fields are not localized; i.e,  $ 1/r^n$ 
vanishes only at $\infty$,  the Zeno effect should prevent collapse 
altogether!  In short, Occam's razor points, inexorably, to rejecting 
the concepts of distributed `particleness,' as well as wave collapse, 
and supports instead admitting the image of concentrated particles at 
distinct locations, which implies that they have preexisting, local 
configuration coordinates---a.k.a.: `hidden variables'---imbedded 
somehow in a separate (pilot) wave aspect.~$^{(11)}$  In the 
imagery supported by SED the wave aspect is engendered by the 
background. 

Hopefully the above inspires an illuminating experiment.

\bigskip\noindent {\bf REFERENCES}\bigskip

\begin{enumerate}

\item R. P. Feynman,  {\it Lectures on Physics III} (Addison-
Wesley, Reading, 1965).

\item In the professional
literature the `peak-and-valley' problem is mostly 
discussed in abstract and abstruse terms only, if at all.  An 
accessible and incisive popular rendition can be found in: D. Wick, 
{\it The Infamous Boundary} (Copernicus Springer, New York, 
1996), p. 53.  

\item See: S. Goldstein, {\it Physics Today} {\bf 51}(4), 38-46; {\bf 
51}(5), 38-42  (1998) and references contained therein for an 
accessible and current account of many alternative interpretations of 
Quantum Mechanics. 

\item See: L. de la Pe\~na and A. M. Cetto, {\it The Quantum 
Dice} (Kluwer Academic, Dordrecht, 1996) for an excellent  
current and very extensive bibliography on and analysis of SED.  

\item H. Puthoff, {\it Phys. Rev. A} {\bf 40}, 4857 (1989) and 
{\it Phys. Rev. A} {\bf 44}, 3385 (1991). 

\item Pilot wave notions can be extended to rationalize the 
Schr\"odinger Equation; see: A. F. Kracklauer, {\it Phys. Essays} {\bf 
5}, 226, (1992). 

\item A. Zeilinger, {\it et al.; Rev. Mod. Phys.} {\bf 60}, 
4, 1067 (1988). 

\item R. Reif, {\it Fundamentals of Statistical and Thermal 
Physics}, (McGraw-Hill, New Yourk, 1965) p. 248.

\item  M. Peshkin and A. Tonomura, {\it The Aharonov-Bohm 
Effect; Lecture Notes in Physics No. 340,}, (Springer, Berlin, 
1989), p. 106. 

\item  F. Hasselbach, in {\it Waves and Particles in Light and 
Matter}, A. van der Merwe and A. Garuccio, eds. (Plenum, New York, 
1994), p. 49.

\item  The hermetic character of Bell's no-go Theorem for hidden 
variables in QM can be doubted. P. Claverie and S. Diner found, 
({\it Israeli J. Chem.} {\bf 19}, 54 (1980)), and this author refound 
({\it New Developments on Foundation Problems in Quantum Physics}, 
M. Ferrero and A. van der Merwe, eds. (Kluwer Academic Pub. 1997) 
p.185) a  mathematical gremlin smack in the middle.  See also 
A. F. Kracklauer, {\it http://xxx.lanl.gov/quant-ph/9812072} for a 
development of this theme.
\end{enumerate}
\end{document}

%% file: neut.tex
\def\dessin#1#2{
\begin{figure}[hbtp]
\begin{center}
\fbox{\begin{picture}(360.00,254.40)
\includegraphics{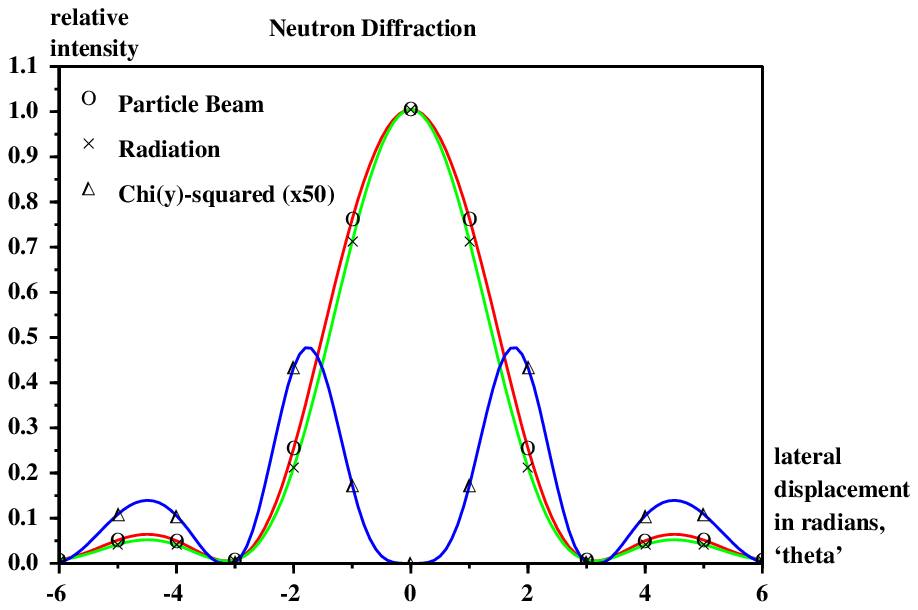}
\end{picture}}
\end{center}
\caption{\label{#2}#1}
\end{figure}}

%% file: electron.tex
\def\dessin#1#2{
\begin{figure}[hbtp]
\begin{center}
\fbox{\begin{picture}(360.00,254.40) 
\includegraphics{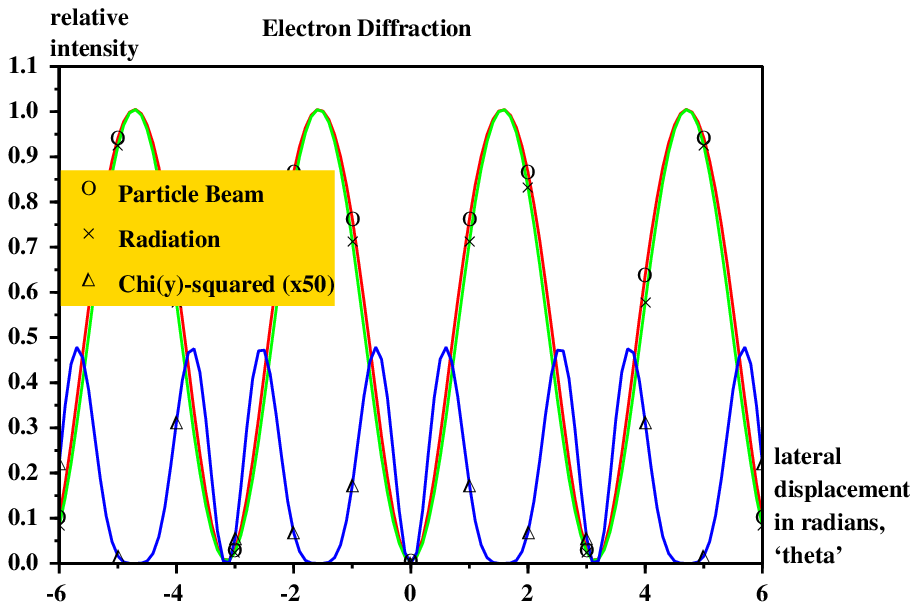}
\end{picture}}
\end{center}
\caption{\label{#2}#1}
\end{figure}}